\documentclass{nature}
\usepackage[table,xcdraw]{xcolor}
\usepackage{graphicx}
\usepackage{amsthm}
\usepackage[labelfont=normal]{subcaption}
\usepackage{amssymb,amsfonts,amsmath}
\usepackage{enumitem}
\usepackage[font=scriptsize]{caption}
\usepackage[labelfont=bf]{caption}
\theoremstyle{definition}

\title{Heterogeneous recovery guided by policies from large scale power failures}
\author{Amir Hossein Afsharinejad$^1$, Chuanyi Ji$^1$ and Robert Wilcox$^2$} 
\begin{document} 
\maketitle
\begin{affiliations}
\item School of Electrical and Computer Engineering, Georgia Institute of Technology, Atlanta, Georgia 30332--0250
\item National Grid, Waltham, Massachusetts 02451
\item Corresponding author: Chuanyi Ji, School of Electrical and Computer Engineering, Georgia Institute of Technology, Atlanta, Georgia 30332--0250, Tel:404-894-2393, jichuanyi@gatech.edu
\end{affiliations}


\section*{Abstract}




Large-scale power failures are induced by nearly all natural disasters from hurricanes to wild fires. A fundamental problem is whether and how recovery guided by government policies is able to meet the challenge of a wide range of disruptions. Prior research on this problem is scant due to lack of sharing large-scale granular data at the operational energy grid, stigma of revealing limitations of services, and complex recovery coupled with policies and customers. As such, both quantification and firsthand information are lacking on capabilities and fundamental limitation of energy services in response to extreme events. Furthermore, government policies that guide recovery are often sidelined by prior study. This work studies the fundamental problem through the lens of recovery guided by two commonly adopted policies. We develop data analysis on unsupervised learning from non-stationary data. The data span failure events, from moderate to extreme, at the operational distribution grid during the past nine years in two service regions at the state of New York and Massachusetts. We show that under the prioritization policy favoring large failures, recovery exhibits a surprising scaling property which counteracts failure scaling on the infrastructure vulnerability.  However, heterogeneous recovery widens with the severity of failure events: large failures that cannot be prioritized increase customer interruption time by 47 folds. And, prolonged small failures dominate the entire temporal evolution of recovery. 

Key words: Recovery guided by policies, power failures, granular data at scale, data analytics

\section*{Introduction}



Natural disasters happen with increasing intensity in the US and the world \cite{NAS17, Marsooli19}. Recovery from disruptions is thus crucial for sustaining our society. In this context, a fundamental problem is whether and how recovery responses guided by government policies is able to meet the challenge of a wide range of disruptions from the moderate to severe and extreme. This problem, while relevant to most infrastructures supporting our society, is particularly acute to the energy grid. Indeed, large-scale power failures were induced by nearly all natural disasters from hurricanes to severe ice storms and wild fire \cite{WH13, NAS17, NAS19}. Millions of people lost electricity supplies for extended durations \cite{NAS17, NAS19}. As such, this work studies the fundamental problem though the lens of recovery from large scale power failures. 



Recovery from large-scale power failures exhibits a typical setting that involves service providers, customers, and government policies which guide the restoration \cite{NAS17, Guikema18, Larsen19}. The complexity of such an interconnection hampers the study of recovery tremendously.  A large volume of work develops algorithms for expediting recovery (see publications \cite{Smith19, Powell19} and references therein). However, field studies on recovery services, i.e., using data from the operational grid and customers, are particularly lacking across a wide range of failure events from the moderate to extreme. As a result, there does not exist an established benchmark measuring the performance of recovery guided by underlying policies \cite{Ji17}. In particular, knowledge is much needed on (a) how the recovery services perform across failure events of different intensity in the first place, and (b) capabilities and fundamental limitations of recovery guided by well adopted policies seen by the field data. 


A key challenge for such a field study originates from the lack of granular data at a large scale. This has taken two forms: a singular focus on individual extreme events, and spatial temporal aggregation. First, failure event data are unevenly adopted across a range of severity; the failure events that have been studied are usually those severe enough to make the headline news \cite{Bloomberg13}. These events are often studied individually, including our own prior work \cite{Wei14, Ji16, Dobson16}. Using individual failure events, however, does not allow a systematic study on whether recovery service can sustain desirable performance from failure events of increasing severity. Second, most work on historic power failures and recovery has used data aggregated over townships and hours since the granular measurements are privately owned and thus difficult to obtain \cite{Dunn19, Ji16}. Advanced data collection and analysis are emerging from smart meter infrastructure and micro PMUs \cite{Yuan19, Meier14}. Unfortunately, such advanced data collection is not widely deployed due to high costs.

This work conducts a large-scale field study of such nature that evaluates recovery services from power failures. Specifically, we focus on recovery services that tie closely with a well adopted by emergency responders in general, i.e., to help as many people as fast as possible under resource constraints \cite{NAS17, FEMA}. Distribution grid operators implement this guiding principle as they triage, i.e., to prioritize restoration of the large failures that affect a high number of customers \cite{FEMA, EEI}. There have been contentious opinions on the services guided by such recovery policy: (a) adopted by providers and disaster responders as a gold standard \cite{FEMA}, but often complained by customers for slow recovery from disruptions \cite{NAS17}, and (b) potential weaknesses demonstrated by a few recent works on algorithm design \cite{Smith19} and case study \cite{Ji16, Roman19}.  There has not been a related systematic study across a wide range of failure events of different intensity. The policy itself has been largely sidelined by the prior work. 

We take a data science approach for the field study. First, we collect large-scale data on recovery from power failures over the years 2011-2019 at the two service regions in the state of New York and Massachusetts. The data are commonly available to, but privately owned by, most of the distribution grid operators in the US and parts of the world \cite{Ji16, Ji17}. The data are granular in geo-location and time, making it possible to relate restoration of the grid to impacted customers at the micro-level \cite{Ji16, Ji17}. Second, we develop a mathematical framework to formulate recovery study through unsupervised learning \cite{Kearns17, Duda00}, an AI approach that is widely used in many application domains but yet to show wide potential in the energy area. The analytics algorithm learns from the non-stationary behaviors of recovery over the failure events of different severity. The data analysis brings new knowledge, centered around heterogeneous recovery in three aspects. 

First, the recovery behaviors are found to be heterogeneous. That is, the failures that affect large and small number of people recover at a disproportional share of customer interruption time.  The large failures, in particular,  affected clusters of customers but were restored with priority for helping the majority of the users. Surprisingly, such disparate recovery counteracts the infrastructure vulnerability found in the prior work \cite{Ji16}, showing the capability of services guided by prioritization policy. 

Second, the disparity, as a generic characteristic of the prioritized recovery, deepens with the increasing severity of failure events. The discrepancy lies in both large and small failures: a significant increase of the large failures that cannot be prioritized, and domination of small failures throughout the dynamic evolution of delayed recovery. These findings tell us that the services governed by the prioritized recovery policy is at the cost of the disparity from both large and small failures; and the cost is significant when failure events become severe and extreme. 

\section*{Results}

\section*{Ensemble of failure events}



We collect the granular data on failure and recovery at the service territories in the two US states during 2011-2019 (see Supplemental Tables 2 and 5). The first data set is obtained from $25,013$ square miles in the state of New York. This data set is used in our primary study of the services guided by the recovery policy. The second data set consists of failure events at the state of Massachusetts over the same time span (see Supplemental Note 6,  Supplemental Tables 4 and 5 for details). The data set from Massachusetts is intended for generalizing our study to a different state. 

The primary data set consists of $53,070$ failures from $172$ events of different severity that affected over $9.7$ million customers. In particular, $60$ such failure events were induced by the so-called major storms declared by the state of New York (see Materials and methods and Supplemental Note 1). Six such failure events are further considered as extreme based on the total number of induced failures and the longest downtime duration of $\sim 170$ hours. For example, Hurricane Irene 2011 and Nor'easters 2018 are two of the named major storms that caused widespread disruptions. The other $54$ failure events that were also caused by major storms are not as extreme and thus referred to as severe failure events. We identify additional $112$ moderate failure events during the same time span. These moderate failures events were not induced by the declared major storms but impacted the grid more than sporadic failures in normal daily operations (Supplemental Note 1). Such moderate failure events were considered as unimportant, and thus have not been included in the prior studies. Together, our data are on the three types of failure events: the moderate, severe and extreme; three types define the severity of failure events. The failure events are induced by a wide range of weather conditions (see Materials and methods). For example, the six extreme failure events were induced by three severe thunderstorms, a hurricane, and two ice storms \cite{NOAA}. 

An overview in Fig. \ref{fig:synopsis_progress} shows the ensemble of failure events of all severities. An individual event has $101 \sim 519$, $106 \sim 1146$ and $1373 \sim 3315$ total failures for the moderate, severe and extreme disruptions, respectively. The recovery services were able to restore electricity service within $24$ hours for every moderate failure event. The downtime duration exceeded $24$ hours per (major) event remained steady for the severe disruptions. The service deteriorated in the extreme failure events, with the downtime up to seven days. Hence, these three groups of disruptions characterize a wide range of failure events of different severity. 


\section*{Distinct categories on recovery speed and failure size} 

\begin{figure*}[h!]
\centering
\begin{subfigure}[b]{0.7\textwidth}
\includegraphics[width=0.9\textwidth]{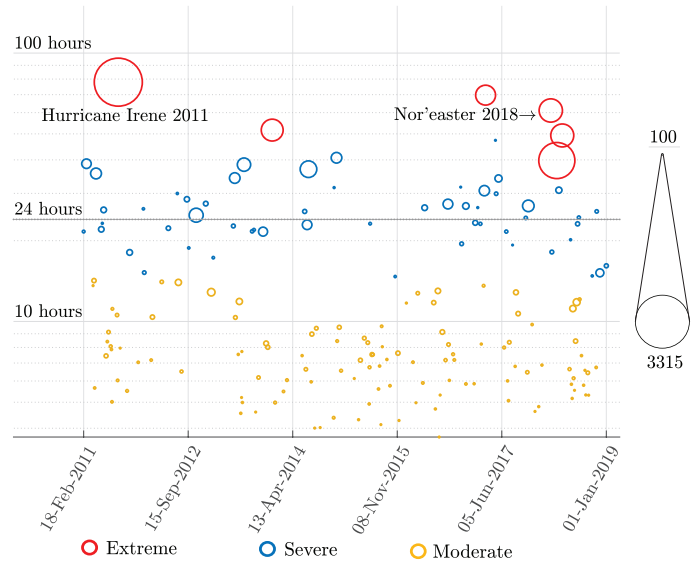}
\caption{\footnotesize Overview of failure events}
\label{fig:synopsis_progress}
\end{subfigure}

\begin{subfigure}[b]{0.32\textwidth}
\centering
\includegraphics[height=4.2cm,width=1\textwidth]{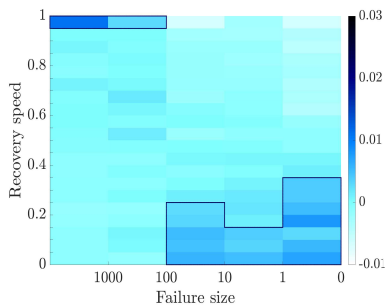}
\caption{\footnotesize Extreme}
\label{fig:inference_C1}
\end{subfigure}
\begin{subfigure}[b]{0.32\textwidth}
\centering
\includegraphics[height=4.2cm,width=1\textwidth]{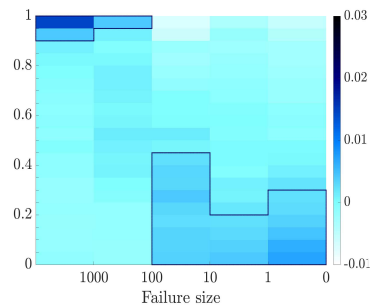}
\caption{\footnotesize Severe}
\label{fig:inference_C2}
\end{subfigure}
\begin{subfigure}[b]{0.32\textwidth}
\centering
\includegraphics[height=4.2cm,width=1\textwidth]{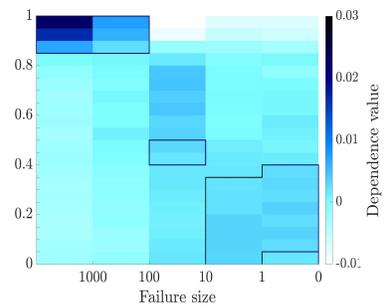}
\caption{\footnotesize Moderate}
\label{fig:inference_C3}
\end{subfigure}

\caption{(a) Overview of failure events: $172$ failure events from the moderate to severe and extreme, February $2011$ to January $2019$. Horizontal axis shows an event occurrence time. Vertical axis is on (the longest) recovery durations in logarithmic scale, where $95\%$ of the failures recovered for each event. The radius of a marker represents the number of failures of a disruptive event from $100$ to $3315$ failures in Hurricane Irene. Red, blue and yellow markers correspond to extreme, severe and moderate events, respectively. (b), (c) and (d): Four categories of recovery speed and failure size. Statistical dependence between the recovery speed and failure size is averaged over the failure events for (b) extreme, (c) severe and (d) moderate disruptions respectively. Horizontal axis shows the failure size and vertical axis represents the recovery speed from $1$ (the fastest) to $0$ (the slowest). Color bar shows the values of statistical dependence. The boundaries highlight the two regions: (i) Prioritized large failures with rapid recovery at the upper left and (ii) Prolonged small failures with slow recovery at the lower right. The average value of statistical dependence within the boundaries exceeds $5\%$ of the maximum value.}
\label{fig:fairness}
\end{figure*} 

We derive a mathematical framework that formulates our study as evaluation of recovery services based on unsupervised learning from the non-stationary data (see Materials and methods). Two covariates emerge as pertinent among the multiple variables relating to our data: (a) the recovery speed as the downtime duration ranked from the shortest to the longest; and (b) the failure size as the number of customers affected by a failure. These two variables together reflect the impact of the recovery policy (i.e., the utility triage) that prioritizes restoration of failures with a high number of affected customers \cite{FEMA}. The unsupervised learning algorithm then obtains the clustered regions in Figures \ref{fig:inference_C1}, \ref{fig:inference_C2} and \ref{fig:inference_C3},  where the recovery speed and failure size are positively dependent (see Materials and methods, and Supplemental Note 2). These clusters partition the space spanned by the recovery speed and failure size into the following four categories for all three types of events: 
\begin{itemize}
\item Prioritized large failures: Each large failure affected more than $100$ customers, and, the recovery speed is ranked within the top $15\%$ shortest downtime durations. Such a fast recovery speed is strongly dependent of the large failure size, showing that the large failures are often restored with priority and rapidly. 
\item Non-prioritized large failures: Each failure has the same large size of affecting more than $100$ customers. The recovery speed, however, is below the top $15\%$ and nearly independent of the failure size. 
\item Prolonged small failures: Each failure affected fewer than $100$ customers, and, the recovery speed is ranked at within the bottom $50 \sim100\%$ of the longest failure durations. The slow recovery speed is strongly dependent of the small failure size, showing that those small failures experienced relatively long downtime durations. 


\item Remaining small failures: Each failure has the same small size of affecting fewer than $100$ customers, and, the recovery speed is ranked higher than the percentage defined by the bottom cluster. The recovery speed is nearly independent of the failure size. 

\end{itemize}

Here, we find that two clustered regions stand out in Figures \ref{fig:inference_C1}, \ref{fig:inference_C2} and \ref{fig:inference_C3}. The first is on the recovery of the prioritized large failures, where the dependence shown by the cluster implies that the large failures are potentially restored with priority soon after experiencing outages. The second is on the recovery of the prolonged small failures, where the long downtime durations are highly dependent of the small failure size, suggesting that the small failures often suffer delays in restoration. Despite the differences in the detailed shapes, these disparate clusters persist from the moderate to severe and extreme failure events. 

\section*{Heterogeneous restoration behavior characterized by recovery scaling law} 

\begin{figure*}[h!]
\centering
\includegraphics[width=0.8\textwidth]{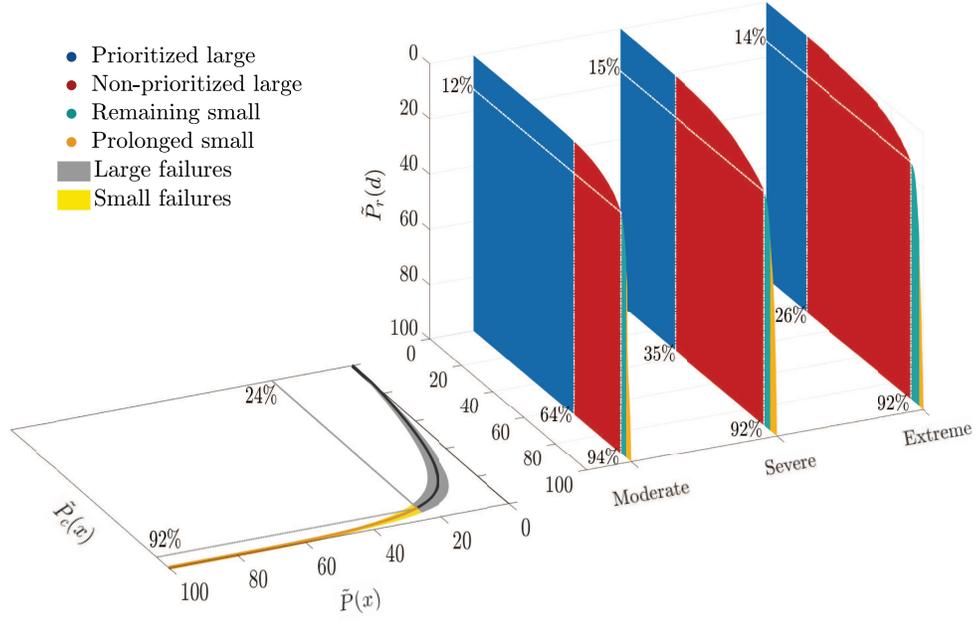}
\caption{Right: Recovery scaling law for the moderate, severe and extreme failure events. $\widetilde {P}_{c}(x)$: the empirical probability as the cumulative percentage of affected customers. $\widetilde{P}_{r}(d)$: the empirical probability $\widetilde{P}_{r}(d)$ as the cumulative percentage of the interruption durations. These probabilities are estimated for the four inference regions: The blue, red, green and yellow colors correspond to prioritized-large, non-prioritized-large, remaining-small and prolonged-small failures, respectively. Left: Failure scaling law. The vertical axis on $\widetilde {P}_{c}(x)$ is the same as recovery scaling, and $\tilde{P}(x)$ is the empirical probability that a disruption affected more than $x$ customers. The failure scaling curve is averaged over all failure events. The shaded regions illustrate standard deviation of the average. The grey and yellow colors are for large (more than $100$ affected customers) and small (fewer than $100$ impacted users) failures, respectively.}
\label{fig:recovery_scaling_law}
\end{figure*}

We define a recovery scaling law to quantify the different restoration behaviors from the four categories. While the failure scaling laws of different forms exist \cite{Clauset09, Ji16}, recovery scaling does not. The recovery scaling law we define is a mapping between two quantities. The first is the probability $P_{r}(d)$ that a failure is restored within the top $d \%$ fastest (ranked by the failure durations from shortest to longest). The second is the probability $P_{c}(x)$ that a customer is affected by such a failure which disrupts the service of more than $x$ users (i.e., $x$ is the failure size). Such a recovery scaling law essentially characterizes relationships between the percentage of customers recovered and cost as the share of interruption durations. The corresponding empirical probability distributions $\widetilde{P}_{r}(d)$ and $\widetilde {P}_{c}(x)$ estimate the scaling property using the data from the inferred clusters (Supplemental Note 3). The resulting empirical recovery scaling law is shown in Fig. \ref{fig:recovery_scaling_law} for three types of disruptive events from the moderate to severe and extreme (Supplemental Note 3). 

The heterogeneous behavior emerges from the recovery scaling law as a prominent attribute. The recovery scaling follows the generalized $90$-$10$ rule, where $92\% \sim 94\%$ affected customers consume $8\% \sim 14\%$ of the total interruption time. Importantly, these $\sim 90\%$ customers are affected by the large failures. Meanwhile, the prolonged small failures, while only amounting to less than $10\%$ of the affected customers, consume $85\% \sim 88\%$ of the total interruption durations. Such disproportionality is present across the moderate, severe and extreme events as shown by the recovery scaling curves of the similar shapes. Phenomenologically, the disparate recovery between the large and small failures is consistent to the typical restoration guided by the recovery policy. The resulting restoration behaviors are bound to be heterogeneous if the recovery policy prioritizes large failures.    

Interestingly, we discover that the recovery scaling mirrors the failure scaling when we extend the failure scaling law in our prior work \cite{Ji16} using the historic data. The failure scaling obeys a generalized $25$-$90$ rule in Fig. \ref{fig:recovery_scaling_law} ($20$-$80$ in layman's terms). A small fraction (top $25\%$) of the large failures is responsible for the majority ($\sim 92\%$) of the affected customers. Basically, a large failure affected a block of customers, which is an infrastructural vulnerability of the distribution grid \cite{Ji16}. The restoration prioritized the large failures so that the blocks of customers recovered in relatively short downtime durations. In comparison, the prolonged small failures amount to a larger portion ($\sim 37\%$) of all failures but affected less than $10\%$ customers. Hence, the data analysis helps connect the two scaling laws in a loop, making us to realize that the recovery services governed by the utility triage effectively counteract the infrastructure vulnerability. 

Additional information on the types of disrupted devices enlightens how the heterogeneous recovery is coupled with the structure of the distribution grid. The prioritized large failures for the severe and extreme events consist of primarily (i.e., $> 89\%$) open substation breakers, with the non-prioritized large failures as reclosers and fused discs (see Supplemental Fig. 3). Substation breakers, located at the main power sources (substations) of the existing distribution grid, have to be restored with priority to provide electricity downstream. Meanwhile, restoring the large failures requires identification and isolation of all failures downstream, for both repair and safety of customers \cite{FEMA}. The time needed for failure identification increases with the scale of severe failure events \cite{FEMA}. In contrast, the prolonged small failures correspond to transformers close to customer property and fused cutouts (see Supplemental Fig. 3 and Supplemental Table 3). This implies that recovery is a complex network problem, where restoration can be dependent across different types of failures, and thus require subsequent study. 
\section*{Disparity deepened in recovery from severe and extreme events} 

The disparity in recovery deepens when the failure events become severe or extreme. The heightened disparity is seen primarily by (a) a significant increased in large failures that cannot be prioritized, and (b) domination of small failures in dynamic evolution of delayed recovery. 

The prioritized large failures consume less than $1\%$ of the total downtime across the failure events of different severity. These prioritized large failures affected on average $1500$ or more customers, each of which recovered in less than $0.1$ hours on average. However, such rapid recovery is difficult to sustain. The percentage of the customers affected by the prioritized large failures is reduced by $29\% \sim 38\%$ from the moderate events to severe and extreme disruptions as shown by the recovery scaling law (see Supplemental Fig. 2 and Supplemental Note 3). In other words, $29\% \sim 38\%$ more customers affected by large failures became non-prioritized shown in Fig. \ref{fig:recovery_scaling_law}. 

\begin{figure*}[h!]
\centering
\begin{subfigure}[b]{0.7\textwidth}
\includegraphics[width=0.8\textwidth]{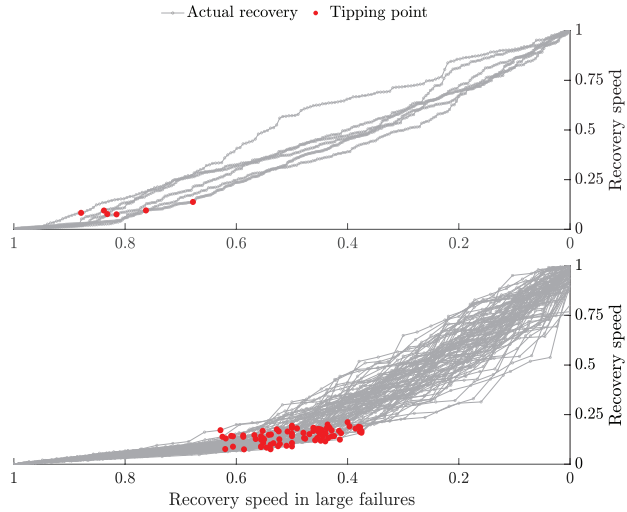}
\caption{\footnotesize Tipping points}
\label{fig:bestefforts-fig}
\end{subfigure}
\begin{subfigure}[b]{0.7\textwidth}
\centering
\includegraphics[width=0.8\textwidth]{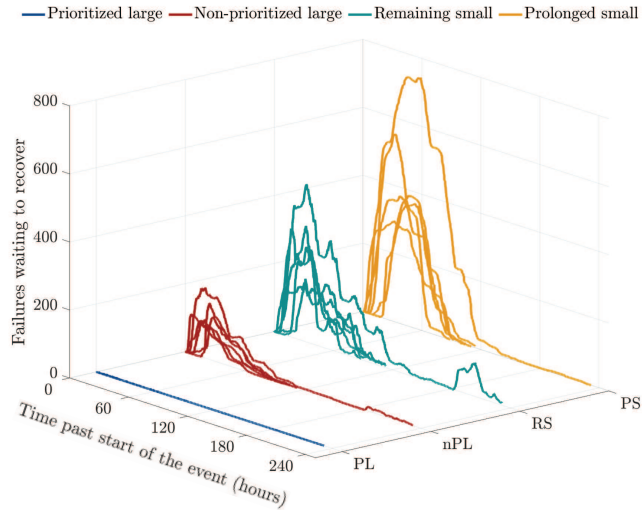}
\caption{\footnotesize Temporal evolution of the four categories of failures }
\label{fig:evolution}
\end{subfigure}
\caption{(a) Tipping points for extreme events (top) and moderate events (bottom). A tipping point corresponds to where the actual recovery speed of large failures starts to deviate from the baseline. Red circles show where the tipping point occurs and the gray curves correspond to the actual recovery for an event. For the  moderate events, the population is plotted for the tipping points with values between $20$-th and $80$-th percentiles for better visualization. (b)Temporal evolution of the four categories of failures during the extreme events. The number of failures waiting to recover at a given time instance is plotted for the prioritized-large (blue), non-prioritized-large (red), remaining-small (green) and  prolonged-small (yellow) failures. Each curve is for one of the six extreme events at the state of New York.}
\label{fig:evolution}
\end{figure*}

To gain further understanding of the non-prioritized large failures, we establish a baseline scenario as if all the large failures were prioritized for restoration (see Supplemental Note 4). The actual recovery speed of the large failures is then compared with the baseline. A tipping point emerges as shown in Fig. \ref{fig:bestefforts-fig}, before which the failures recovered consistently fall in the prioritized large category (see Supplemental Fig. 5 and Supplemental Note 5). The average tipping point values shown in Fig. \ref{fig:bestefforts-fig} reduce from $49\%$ for moderate events down to $26\%$ then $19\%$ respectively for severe and extreme disruptions (Supplemental Note 5). Here, the $30\%$ decrease in the level of prioritization of the large failures matches the $38\%$ reduction of the rapidly recovered customers in the recovery scaling curves from the moderate to extreme disruptions in Fig. \ref{fig:recovery_scaling_law}. Therefore, the disparity deepens in the recovery of the prioritized and non-prioritized failures when a disruptive event becomes severe or extreme. 

Dynamic evolution of recovery from extreme events further shows vividly in Fig. \ref{fig:evolution}(b), how the large and small failures recover differently throughout an entire disruption (Materials and methods, and Supplemental Fig. 8 for non-stationary failure-recovery processes).  First of all, at a given time during a disruption, there were more prolonged small failures waiting for restoration than any other categories. This shows that the prolonged small failures dominate delayed recovery. Next, the non-prioritized large failures that were pending for repair reach the maximum value about an average of $24$ hours earlier than the prolonged small failures. Afterwards, the non-prioritized large failures enter the recovery stage (Materials and methods), and recover at a faster rate than the prolonged small failures. Comparing the worst impact at the peak, the maximum value is three times larger on average for the prolonged small failures than that of the non-prioritized large ones. In comparison, the prioritized large failures have the fewest pending repairs at any given time across all the extreme events. And, the the remaining small failures exhibit diverse variations. Therefore, small failures dominate delayed recovery during the extreme events. 

How significant is the deepened disparity? The impact on customers provides an answer. First, the non-prioritized large failures result in the largest average customer interruption time (CMI - Customer Minute of Interruption) among the four failure categories. Importantly, the $30\%$ increase of the non-prioritized failures results in $47$ times more average CMI from the extreme disruptions than that of the moderate events (Supplemental Note 6). Next, our analysis finds that the growth rates of the actual cumulative customer downtime from the extreme events are drastically different for the four categories of failures (see Supplemental Fig. 6). The prolonged small failures stand out whose total downtime soars at a fastest rate, $3.5 \sim 5.7$ times of that for the non-prioritized large failures. Furthermore, the prolonged small failures affected $60 \sim 800$ customers for a moderate disruption but $3500 \sim 8700$ users for an extreme event. The customers affected by the prolonged small failures regain power within $24$ hours for the moderate events but experience interruptions as long as $170$ hours for the severe and extreme disruptions. There is also $20\%$ increase of the smallest failures that affected one customer from the extreme disruptions compared with the moderate events (also observed by the Nor'easters in New Jersey 2018 \cite{NJ18}). About $48\%$ (i.e., $1444$) of those smallest failures in an extreme event are prolonged on average, representing the longest interruption time. This demonstrates that the extreme events induced particularly more failures close to customer properties, which are likely to experience delayed recovery. Overall, the impact is significant when the disparity deepens, reflected by the impact on customers from both large and small failures demonstrated across the service region at New York state in Fig.4. 


\begin{figure*}[h!]
\centering
\begin{subfigure}[t]{0.7\textwidth}
\centering
\includegraphics[width=0.8\textwidth]{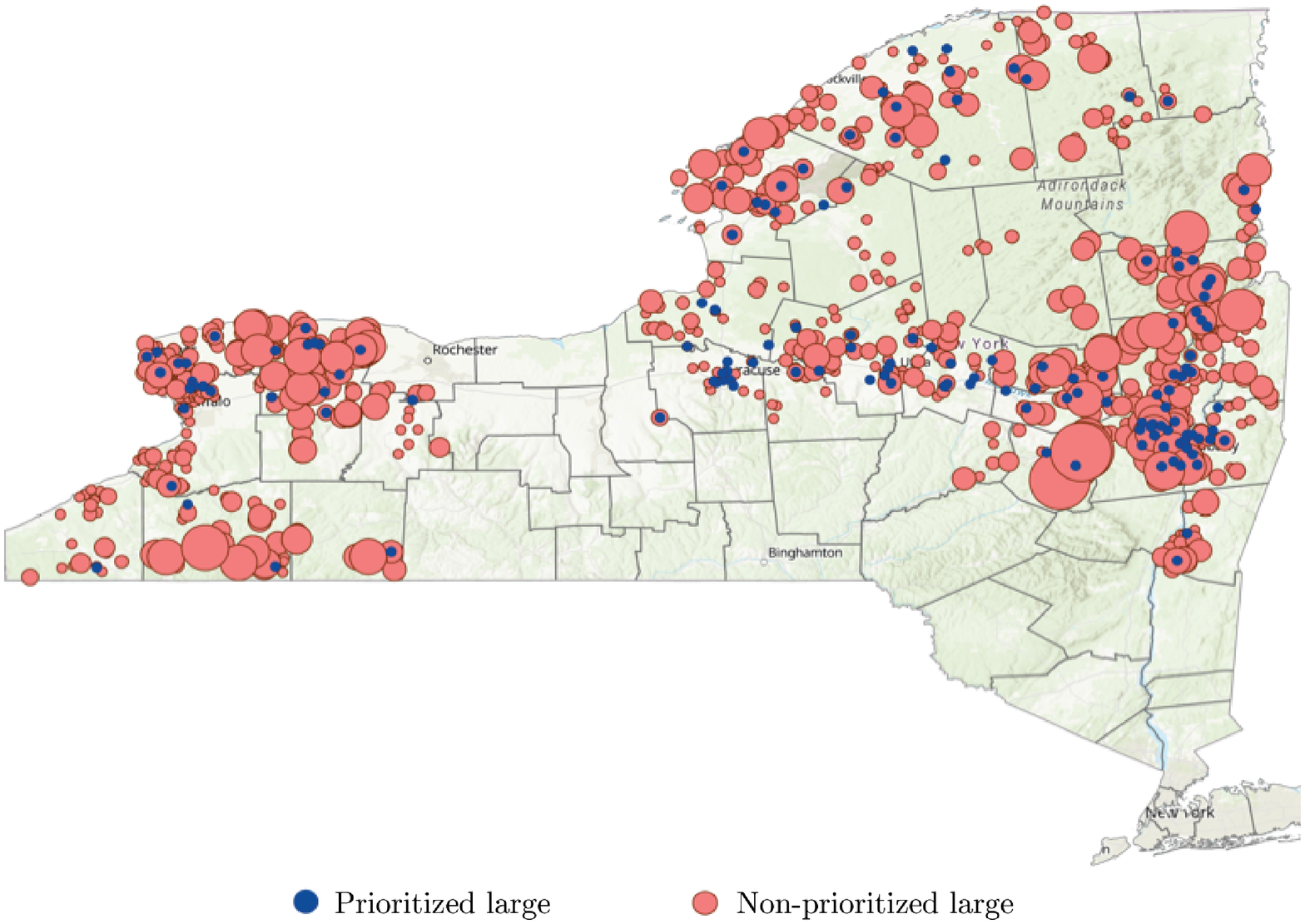}
\caption{}
\label{fig:geo_large}
\end{subfigure}
\begin{subfigure}[t]{0.7\textwidth}
\centering
\includegraphics[width=0.8\textwidth]{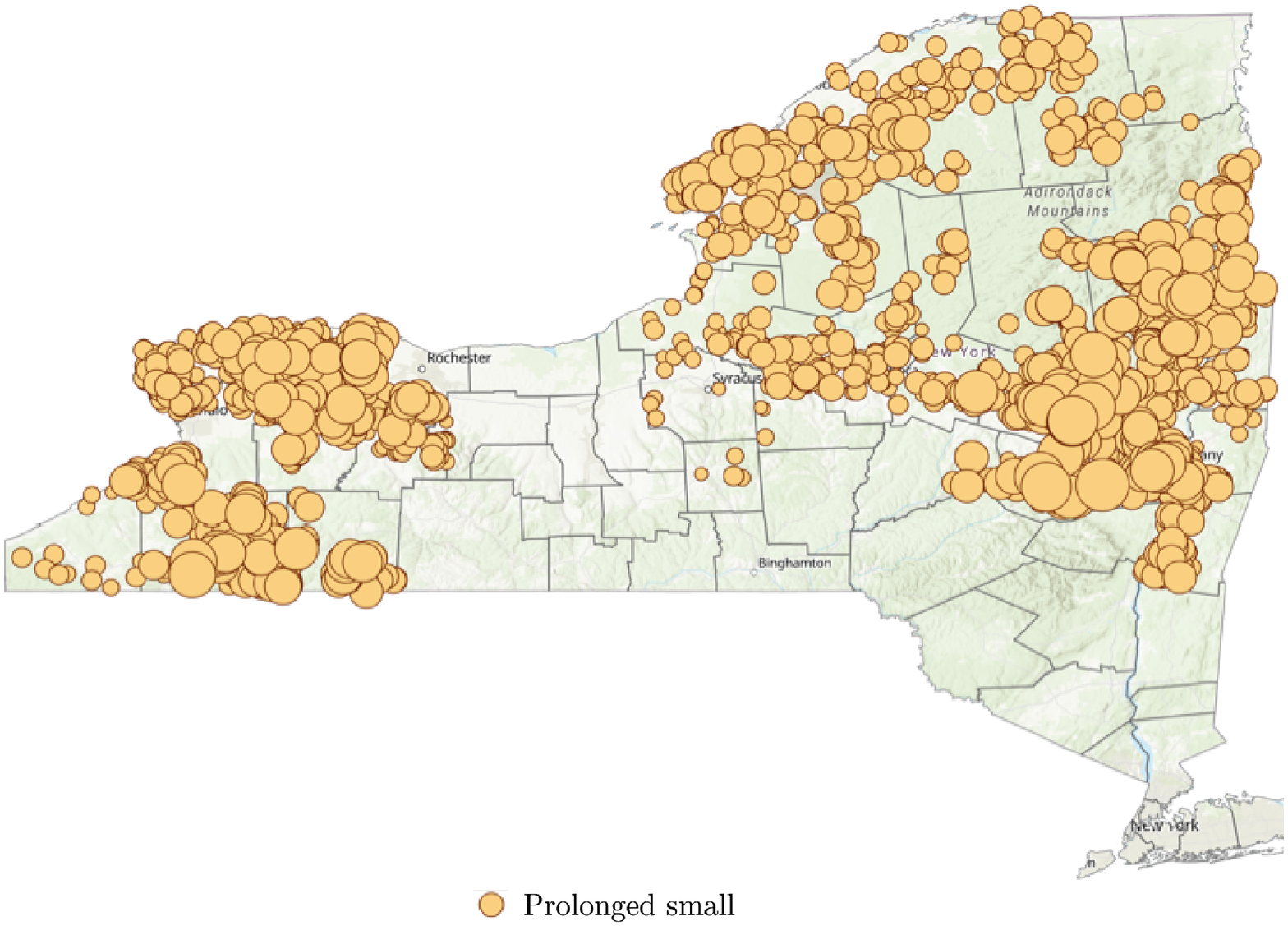}
\caption{}
\label{fig:geo_prolonged_small}
\end{subfigure}
\caption{Geo-location of (a) large and (b) prolonged small failures in extreme events. The radius of the circles correspond to downtime durations from $0$ to $150$ hours. The blue markers in (a) correspond to the prioritized large failures. The red markers in (a) show the remaining large failures. The yellow markers in (b) represent the prolonged small failures in the extreme events.}
\label{fig:geo_extreme}
\end{figure*}

\section*{Generalization to the state of Massachusetts} 
Can our analysis scale up to different service territories? Different service regions are known to have different grid structure and weather conditions \cite{NAS17}. However, minimizing the impact of failures on customers under constraints is a principle of recovery that guides restoration services in all regions by FEMA \cite{FEMA}. Thus, it is important for us to explore the generalizability of our approaches to entirely different locations that are nevertheless under the same recovery guideline. We conduct a case study, focusing on the recovery from a major event, Super Storm Sandy, by the same service provider but in the state of Massachusetts. 


The state of Massachusetts has fewer than $10$ major storms from $2011$ to February of $2019$ at the service region in Massachusetts (see Supplemental Note 5). Super Storm Sandy in 2012 is one of the major storms that caused $4,011$ failures and affected $330,298$ customers for up to five days. Furthermore, the service region in Massachusetts is smaller (i.e., $\sim 3,870$ square miles) but more densely populated than that at the state of New York. 


We obtain the inference diagram (see Supplemental Fig. 5a) from the data at Massachusetts, where the clustered regions are identified using the same parameters as those in Supplemental Note 2. The clusters do emerge for the prioritized-large and prolonged small failures, albeit the shapes of the clusters are different from those of New York State. We further obtain the recovery scaling law for this event in Supplemental Fig. 5b. Compared with that for New York State, the scaling curve has a similar shape and follows the similar $90 \sim 10$ rule. The prioritized-large and prolonged-small failures also exhibit the outstanding disparity in the percentage of affected customers and consumed downtime. The actual impact includes $338$ large failures in total (and $35$ prioritized large failures) that affected $42,554$ customers with $39$ hours aggregated downtime. As a comparison, the prolonged small failures are four times as many but correspond to $\frac{1}{5}$ fraction of the affected customers with $2,059$ times longer total downtime hours. These results demonstrate both the promise for scaling up our analysis to other US states, and the need of incorporating differences in the grid structure and regional policies.

\section*{Discussions}

This work performs a large scale field study on the recovery from large scale power failures using granular data from the operational distribution grid across a wide range of disruptions in two US states. Guided by unsupervised learning from non-stationary data, our analysis has drawn new and insightful knowledge on the recovery services governed by the commonly adopted government policy. 



The restoration-service guided by the prioritized recovery policy is found to counteract the infrastructural vulnerability of the power distribution grid. Phenomenologically,  the recovery practice, quantified by a novel recovery scaling law, prioritizes the restoration of the large failures so as to reinstall services for the majority of affected users. The resulting performance persists across the failure events of different severity, showing a desirable property of recovery. 


The prioritized recovery, by design, does not optimize rapid restoration of all failures. In fact, heterogenous recovery heightened with the intensity of failure events for both large and small failures. There was a significant increase of the non-prioritized large failures that resulted in the largest total customer interruption time. This shows that the rapid restoration from prioritization does not sustain to severe and extreme failure events. Meanwhile, at any given instance during an extreme event, there were statistically more prolonged small failures waiting for repair than the large ones. Further, the prolonged small failures affected thousands of customers in an extreme event, each of whom experienced days of interruption. Hence, the impact of the disparity from severe and extreme events is significant. And, the depended disparity shows a fundamental limitation of the recovery under the prioritization policy. 

Everyone who suffers from a natural disaster is important to the recovery services \cite{Roman19}. A research question is how to overcome the fundamental limitation of the recovery. Our data analysis tells us that the restoration is constrained additionally by the structure of the distribution grid, where a certain large failures relating to the power components upstream have to be restored first \cite{Wei14, Ji16}. Hence, a combined enhancement on the infrastructure \cite{Dey19}, recovery policy and disaster preparation is clearly a necessity, in order to reduce the interruption time to all users \cite{Larsen19, Gorham19, Baik20}. In this context, our study provides a benchmark for comparing an enhancement at the magnitude and scale of an operational grid. 



The data we use in this work are commonly available to most distribution grid operators in the US and parts of the world \cite{Dunn19, Ji16, Ji17}. Thus, this study demonstrates that energy service providers have the ability to adopt data science \cite{Donoho17, Blei17, Yu20}, specifically, to turn their own data into knowledge, to benchmark and improve recovery. Further, such methodology adoption may offer profound impact to the community. Traditionally, a stigma runs deep that data-driven study on recovery may potentially reveal limitations of services. In fact, it is simply not a common practice for providers to study and share their own performance of service interruption using their own data.  We hope that this study will encourage service providers and policy makers to take an active role in data analytics, to nurture a more collaborative environment. 




\section*{Experimental procedures and methods}

{\bf Unsupervised learning of recovery behavior from data:} We derive a mathematical formulation for obtaining new knowledge on recovery services from the data. This is motivated by a formulation from Kearns et.al that evaluates the fairness of policies in machine learning \cite{Kearns17}. Here we view our study as evaluating the performance of recovery with the variables on failure characteristics and the restoration policy. The performance of recovery is not known. Neither is the relationship between the recovery performance and failure characteristics. Therefore, unsupervised learning is best suited for inferring new knowledge from the recovery behaviors and failure characteristics. 

We let $X \in R^n$ represent failure characteristics where $n \ge 1$. The failure size is one such characteristic, which is the number of customers affected by a failure. The type of disrupted devices is another failure characteristic, which includes substations as the primary energy sources at the distribution grid, and transformers close to customer property \cite{Ji16, Ji17}. Both the failure size and type of disrupted devices characterize the infrastructure of the distribution grid. $X \in A$ represents a set ($A$) of values for the variables (e.g., $100$ or more affected customers, and damaged transformers at certain geographical coordinates). We let $Y$ ($0 \le Y \le 1$) represent the recovery behavior on how fast a failure recovers, where $Y$ is obtained by ranking the downtime duration from the shortest to the longest. Thus, $Y \in B$ represents the recovery speed, which is slow when $B: 0 \le Y \le y_1$, or rapid when $y_1< Y<1$. And, the boundary between the slow and rapid recovery is chosen as $y_1=0.5$ for simplicity. Other feature variables exist (e.g., weather and terrain conditions) that vary from event to event, and thus are considered as random factors. We let $Y$ be a random variable characterizing the behaviors of recovery. 
 
The recovery speed ($Y$) and failure characteristics ($X$) are random variables that in general are statistically dependent \cite{Wei14, Ji16, Ji17}. The statistical dependence of $(X,Y)$ is represented by their joint probability distribution $P(X \in A, Y \in B)$. For example, the strong dependence between the recovery speed and failure size reflects the impact of the utility triage on restoration. The independence of the recovery speed and failure characteristics provides a baseline ($P(X \in A) P(Y \in B)$). A metric ($f_{A,B}$) measures the degree of the dependence through a joint probability distribution relative to the baseline, where 

\begin{equation}
f_{A,B}=P(X \in A, Y \in B) - P(X \in A) P(Y \in B). 
\end{equation}

The recovery speed and failure characteristics are positively dependent when the metric exceeds a threshold. 

The unsupervised learning is implemented by first estimating the empirical joint probability $\tilde{P}(X \in A, Y \in B)$ using our data. Then the maximum coverage regions on $(A,B)$ are obtained for $f_{A,B}$ to exceed a chosen threshold of $5\%$ of the maximum value. And, such a condition is satisfied with a sufficiently small error bar $Err(f_{A,B})$. The error bars are obtained through $5$-fold cross validation for each event and averaged between the training and test data across a given type of failure events (see Supplemental Note 2). The resulting coverage regions correspond to clusters in unsupervised learning \cite{Duda00}. Such an unsupervised learning algorithm is applied to the three types of the failure events from the moderate to severe and extreme (see Supplemental Note 2). 

{\bf Non-stationary recovery of the four categories: } Failure and recovery are modeled as non-stationary random processes in our prior work \cite{Ji16, Ji17}. The failures waiting to recover, referred to as pending repairs, represent the first-order interaction of failure and recovery processes \cite{Ji16, Gallager14, Hajek15, Bertsimas97}. The maximum number of the pending repairs separates the failure-recovery processes into two stages, as illustrated by a typical example of an extreme failure event in Supplemental Fig. 8.  While the failures can occur and recover at any time, the first stage is dominated primarily by the failure occurrences until the number of pending repairs reaches the maximum value. The second stage is dominated by the recovery process, where recovery rate exceeds the failure rate, reducing the number of pending repairs from the peak. The slope for the number of pending failures to decrease from the peak value illustrates the rapidity of restoration. 

Such non-stationary properties are extended directly to include the four categories, showing the dynamic evolution of delayed recovery. Let $R(t |c_i)$ be either the number of failures (or affected customers) in Category $c_i$ ($1 \le i \le 4$), waiting to recover at time $t$. Then 
\begin{equation}
R(t |c_i) =  E \{\int_0^t G(t,v|c_i) dv\},
\end{equation}
where $G(t,v | c_i)$ is either a failure or customer affected in Category $c_i$, $1 \le i \le 4$, that occurs at time $v$ but yet to recover at time $t$, $0< v < t$. The integration adds up over all such failures or affected customers occurred in $[v,t]$. The expectation $E$ is over randomly occurring failures or affected customers \cite{Ji16, Ji17}. $R(t |c_i)$ is estimated from the data on the four categories of failures (or affected customers) waiting to recover. 


{\bf {Data and temporal processing:}} The non-stationary property is also used for selecting a subset of the data samples for reliable analysis. Two sets of data are collected by this work. One is the primary data set for learning and analyzing restoration behaviors governed by a commonly-used prioritization policy on recovery. Such a data set is collected from a service region in the state of New York. The other is for extending analytics to the state of Massachusetts, particularly on the impact of different policies for declaring major failures events across state borders.  We use the same variables for both data sets, and thus focus on describing the details of the primary data. 

The main data set is obtained from a service region of $25,013$ square miles at the state of New York. The data set includes $135,999$ failures collected by a distribution system operator (DSO) from $2011$ through January $2019$. Total $33,497,547$ customers were affected, with customer downtime duration of $778,718$ hours (see Supplemental Table 2). 

Each data sample contains detailed information regarding failures and recoveries: failure occurrence and recovery time in minutes, the number of affected customers per failure, the device type, geo- and grid-location of the disrupted components. Here a failure is represented by a damaged power component such as a transformer, or an activated protected device from an open substation breaker to a blown fuse. The grid locations of failures are highly coupled with the device types \cite{Ji16, Ji17} and thus not used in this work. Geo-locations are used mostly for visualization. 

This work focuses on the available data to distribution system operators where failures were reported by customers. As such, some of the small failures might not be reported on time until discovered later. Some other outages, particularly those during recovery process, may be induced by restoration itself that had to turn off the electricity supply for repair. Thus, we extract reliable data for our study based on the non-stationary random processes discussed above \cite{Ji16}. To be specific, most failures with long durations occurred at the first stage when restoration was overwhelmed by rapidly occurring failures. Hence, the maximum number of pending repairs is an indicator for separating data samples into the two stages for severe failure events (see Supplemental Note 7). 

A total of $66\%$ of failures in extreme events occurred in Stage $1$ while the remaining $34\%$ at Stage $2$. In particular, at the failure stage (i.e., Stage $1$), there are $87\%$ of large failures (each of which affected $100$ or more customers); $75\%$ of small-size failures except for one customer failures (each of which affected $2 \sim100$ customers. For the failures where each affected one customer, $53\%$ occurred in Stage $1$ whereas the remaining were in Stage $2$. Therefore, given the significantly larger impact, both in terms of the number of customers affected and interruption durations, we use failures that occurred at the failure stage of the extreme and severe disruptions for our analysis. 

We note that the majority of the failures that occurred during the recovery stage has relatively short interruption  durations. Some of those interruptions on electricity supplies were indeed planned outages for necessitating repairs. Hence, the failures occurred at the first stage represent the majority induced by exogenous weather events (consistent to the data we used in the prior work \cite{Ji16}). Additionally, the failure data at Stage $1$ are considered to be reasonably accurate in interruption durations, since certain small failures were found during the recovery stage which might have occurred earlier (this of course can also occur to the small failures at Stage $1$ to a lesser degree). Therefore, we use the data from the failure stage (Stage $1$) in our analysis. Overall, using the data from Stage $1$ provides conservative estimates of recovery behavior, since our data preprocessing extracts failures that were naturally initiated by an exogenous event (consistent to the data we used in the prior work \cite{Ji16}). Different from what used in our prior work \cite{Ji16}, the data we leave out in Stage 2 exclude a minute portion of failures with short durations. As such, our results here are more conservative in estimating the impact of the small failures than the prior work \cite{Ji16}. While this may introduce a minute difference in actual percentages of small failures, the findings are consistent through out our previous and current study: There were more (prolonged) small failures than the large ones, and those small failures affected a lot fewer customers but experienced much longer downtime durations. 

{\bf Limitations of data and methods:} Data used by this work enable learning the behavior of recovery, which results in insights on the policies. However, additional data are needed on terene conditions, topological characteristics of the grid, and detailed impact on customers. Such data will help identify causal relations beyond behavioral learning, which is needed for enhancing services and policies as well as infrastructure enhancement. Further, additional micro data are needed from more US states. This will allow us to scale up our analysis and insights to more service territories. 



\section*{Acknowledgements}

The authors thank Yun Wei, Scott Ganz, Richard Simmon, Mark Davenport for helpful discussions, Ramachandra Sivakumar for providing ArcGiS for data visualization, Eric O. Hsieh and Byeolyi Han for helping with data processing. Partial support from Strategic Energy Institute at Georgia Tech is gratefully acknowledged. 








\end{document}